# Plasmonic nanogap enhanced phase change devices with dual electrical-optical functionality

*Nikolaos Farmakidis[1][†], Nathan Youngblood[1][†], Xuan Li[1], James Tan[1], Jacob L. Swett[1], Zengguang Cheng[1], David C Wright[2], Wolfram HP Pernice[3] and Harish Bhaskaran\*[1]*

[1]Department of Materials, University of Oxford, UK
[2]University of Exeter, UK
[3]University of Munster, Germany

[†]These authors contributed equally.

\*Corresponding authors. E-mail: *harish.bhaskaran@materials.ox.ac.uk*

Keywords: Integrated electro-optical memories, phase change photonics, Mixed-mode PCM

**Abstract**

Modern-day computers use electrical signaling for processing and storing data which is bandwidth limited and power-hungry. These limitations are bypassed in the field of communications, where optical signaling is the norm. To exploit optical signaling in computing, however, new on-chip devices that work seamlessly in both electrical and optical domains are needed. Phase change devices can in principle provide such functionality, but doing so in a single device has proved elusive due to conflicting requirements of size-limited electrical switching and diffraction-limited photonic devices. Here, we combine plasmonics, photonics and electronics to deliver a novel integrated phase-change memory and computing cell that can be electrically or optically switched between binary or multilevel states, and read-out in either mode, thus merging computing and communications technologies.

**Introduction**

Although integrated photonics has gained great traction over the last decade, primarily for their potential to overcome fundamental limitations of today's electronic circuitry (*1*, *2*), the conversion of electrical and optical signals seamlessly on a chip has remained elusive. The development of compact devices for efficient electro-optic conversion holds great importance as sharing the computing load between the electrical and optical domains shows increasing promise for applications including integrated optical switches, reconfigurable photonic circuits, photonic artificial neural networks, and more (*3–5*). Phase change materials are considered outstanding candidates for dual mode operation as they in principle provide both electrical and optical modulation functionality. To this effect, several devices implementing non-volatile, optical phase-change materials (PCMs) have been proposed (*6–8*), but none have been successfully demonstrated on an integrated platform. This is because the high electrical contrast between the conductive and insulating state in PCMs requires very close spacing between the metal contacts (usually tens of nm) to initiate a phase transition (*9*). Additionally, the resulting conductive region formed after electrical switching is at most a few hundred nanometres in diameter—thus reducing the total volume of material for light-matter interaction (*10*).

Combining plasmonics with PCMs is a particularly promising approach for satisfying such stringent requirements, since the dimensions of such devices can be reduced to tens of nanometres and smaller—significantly below the diffraction limit of traditional optical devices (*11*, *12*). The combination of high electrical conductivity and strong plasmonic resonance at optical wavelengths in silver and gold has led to extremely compact electro-optic nanogap devices such as integrated light sources (*13*), photodetectors (*14*, *15*), and modulators (*16*, *17*). Additionally, the extremely high field-enhancement possible with sub-wavelength nanogaps enables very high-sensitivity spectral measurements for applications such as label-free detection of biomolecules (*18–20*).

While plasmonics permits very strong light-matter interaction in nanometre-scale devices, the relatively high loss of metals at optical frequencies makes guiding light inefficient. Combining integrated photonics with nanoscale plasmonics however allows for both low-loss light delivery and strong light-matter interaction in a compact footprint (*18, 21*). In this paper, we combine waveguide-integrated plasmonic nanogaps with a phase-change material, $Ge_2Sb_2Te_5$ (GST), to create an electro-optic memory cell that is fully addressable in both electrical and optical domains. Previous demonstrations of such mixed-mode devices have either used non-volatile phase-change materials, such as $VO_2$ (*22, 23*), which requires significant power consumption for data retention, or have been limited to write/erase operations either electrically or optically, but not both (*24–26*). By exploiting both the nanoscale dimensions and strong field confinement of a plasmonic nanogap, we enable both electrical and optical non-volatile switching of GST within the gap allowing for full mixed-mode operation of a PCM memory cell.

**Results and Discussion**

A 3D illustration of our device can be seen in Fig. 1a. We use a partially-etched $Si_3N_4$ rib waveguide to route the optical signal to the plasmonic memory cell which are coupled via a tapered geometry (*27–29*). Plasmonic nanogaps are formed between two metal electrodes (3nm Cr / 75 nm Au) fabricated via lift-off using electron-beam lithography and thermal evaporation. A thin film (75 nm) of GST with a 5 nm capping layer of $SiO_2$ bridges the nanogap, controlling both the electrical resistance and optical transmission of the device depending on the state of the material. By sending either electrical or optical pulses, we can reversibly switch the GST within the nanogap between its highly resistive amorphous phase and conductive crystalline phase (*30–32*). Optical and SEM micrographs of the completed device are shown in Fig. 1b-d.

In order to quantify the field enhancement of the plasmonic nanogap, we performed both 2D eigenmode and 3D finite-difference time-domain simulations (FDTD) using Lumerical Solutions® and plot the field profile cross-sections of the device when GST is in the amorphous and crystalline phases (see Fig. 1e). The amplitude of the electric field intensity is scaled relative to the field amplitude of the waveguide mode. Fig. 1e shows that the electric field intensity is enhanced by more than an order of magnitude in the case of amorphous GST due to strong field confinement within the 50 nm nanogap. This enhancement reduces by a factor of 5 when the GST within the nanogap switches to the strongly absorbing, crystalline state. Fig. 1f shows the simulated transmission spectrum for the complete waveguide-nanogap system. Here we observe that at longer wavelengths, the transmission actually increases when GST in the nanogap is crystalline. This is a result of the modulation of the coupling between the nanogap and waveguide where an increase in the wavelength-dependent refractive index causes an increased coupling to the plasmonic mode within the nanogap. Indeed, we see in 3D FDTD simulations that the light which is scattered and reflected between the waveguide/nanogap interface decreases when the GST is in the crystalline state. Although the optical absorption also increases for crystalline GST, the enhanced coupling to the nanogap results in an overall increased transmission when GST is switched from the amorphous to the crystalline state.

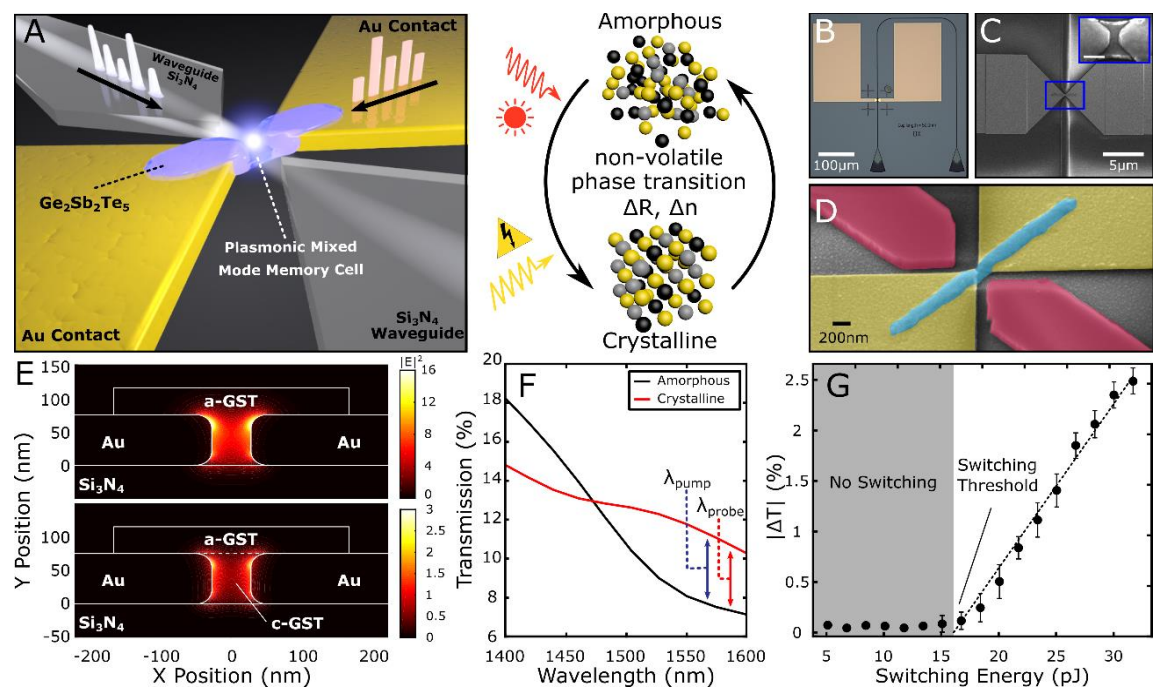

**Figure 1:** Mixed-mode plasmonic memory cell integrated in a photonic waveguide. **a)** 3D illustration of device concept. Light is delivered to the nanoscale device via a photonic waveguide, while the Au contacts serve as both device electrodes and plasmonic nanogap to focus incoming light. **b)** Optical and **c-d)** SEM images of device after fabrication (scale bar is 100 nm in inset of **c)**). The width of the nanogap was measured to be approximately 50 nm for the devices used. **e)** Eigenmode simulations of field enhancement inside the plasmonic nanogap when the GST is in the amorphous (top plot) or crystalline state (region between Au electrodes in lower plot). The field enhancement is much stronger when GST is in the amorphous state due to the significantly lower optical loss. **f)** FDTD simulation of the transmission of device before and after crystallization. The significant change in the refractive index changes the coupling between the nanogap and waveguide which reduces reflection from the input waveguide, thereby increasing overall transmission of the device in the crystalline state. **g)** Experimental measurement of total energy in the waveguide required to achieve a non-volatile phase transition. The switching threshold was measured to be 16 ± 2 pJ according to a linear fit to the data (black dashed line).

We experimentally verify this plasmonic field enhancement by sending pulses of increasing amplitude to the device and measuring the non-volatile change in transmission of a counter-propagating probe signal ($\lambda$ = 1570 nm). The measured change in transmission as a function of pulse energy can be seen in Fig. 1g. Here we observe a change in transmission with a pulse energy of 16 ± 2 pJ using a 5 ns optical pulse. We note that this energy is significantly lower than previous demonstrations of evanescently coupled phase-change devices (*30*) due to the strong field enhancement and small mode volume of our plasmonic nanogap.

We then perform optical programming of our phase-change memory cell; we send optical write and erase pulses to partially amorphize and crystallize the GST within the nanogap while monitoring both the optical transmission and electrical resistance. Fig. 2a shows a schematic of the experimental setup. Piece-wise optical write pulses (7.5 mW for 8 ns followed by 3 mW for 400 ns) and rectangular erase pulses (7.5 mW for 8 ns) are used to switch the GST between crystalline and amorphous states respectively. We use a constant-power optical probe to monitor the change in transmission, while a source-meter unit (SMU) in constant voltage mode ($V_{bias}$ = 50 mV) is used to monitor the change in resistance. Time-dependent traces of the simultaneous change in both the transmission and resistance of the device can be seen in Fig. 2b during consecutive optical write and erase pulses separated by 1 second. In agreement with our FDTD simulations of Fig. 1f, the resistance and transmission traces change as expected – i.e. an amorphization (erase) pulse results in an increase in electrical resistance and decrease in optical transmission while a crystallization (write) pulse results in the opposite effect.

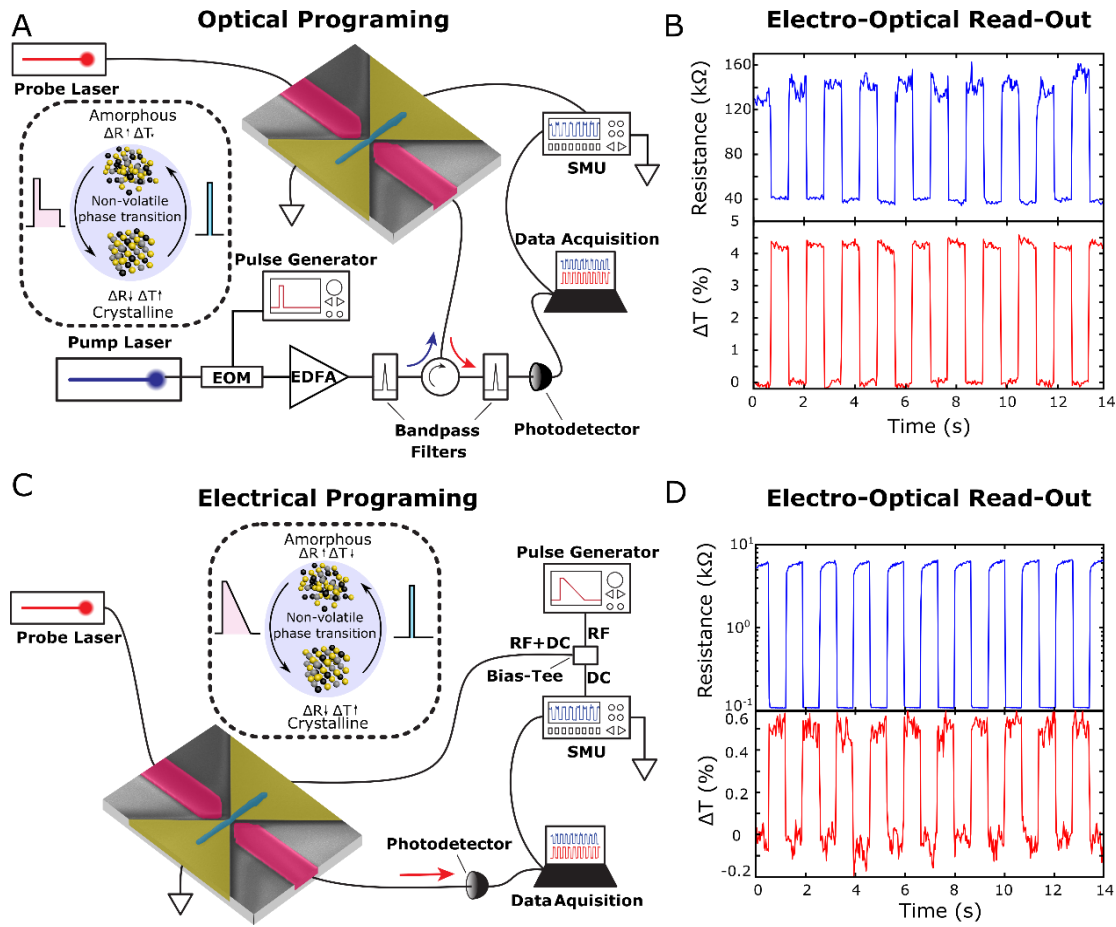

**Figure 2:** Non-volatile electro-optical programming of mixed-mode plasmonic memory. **a)** Illustration of the experimental setup used to measure optical programming of mixed-mode device. Optical write (piece-wise pulse: *7.5 mW for 8 ns + 3 mW for 400 ns*) and erase (*7.5 mW for 8 ns*) pulses are used to switch the GST between crystalline and amorphous states respectively. A CW optical probe signal and constant voltage source are used to monitor the optical transmission and electrical resistance of the GST simultaneously. **b)** Mixed-mode measurement during sequential optical write and erase pulses. Both the optical transmission and electrical resistance can be used to monitor the state of the GST within the nanogap. **c)** Illustration of experimental setup used for electrical programming of device. Electrical write (rectangular: *350 mV for 10 ns*) and erase (triangular: *350 mV with 5 ns/500 ns rise-fall time,*) pulses are used to switch the state of the GST. Again, a CW optical probe signal and constant voltage source are used to monitor the transmission and resistance simultaneously. **d)** Mixed-mode read-out of the device's transmission and resistance state.

We subsequently demonstrate successful operation of the device in the electro-optic domain, wherein a change in optical transmission is observed as a result of electrical switching of the device. Here, we add a bias-tee between the device and SMU to monitor the DC resistance of the device while sending write and erase pulses via the RF port of the bias-tee (see Fig. 2c). By sending a 10 ns, 350 mV pulse across the device (5 ns rise and fall time), we amorphize the GST in the gap, while a 350 mV triangular pulse (5 ns rise, 500 ns fall time) recrystallizes the GST. As shown in Fig. 2d, the state of the device can be seen in both the optical transmission and electrical resistance read-out. As we observed previously and as expected from our FDTD models, the transmission increases when the GST is switched to the crystalline state (see Fig. 2b) and is repeatable for many cycles as shown in Fig. 3d and e. However, in these measurements, we observe higher contrast in the electrical resistance than during optical switching. This is because, we are able to switch only the volume of material needed to create or disrupt a conductive path between the electrodes (*33*). The enhanced level of optical sensitivity to such a small volume of material switching between its amorphous and crystalline phase is attributable to the strong light-matter interaction within the plasmonic nanogap. Crucially, the voltage required to switch the state of the GST is aided by reducing the spacing between the metal contacts and thereby increasing the electric field within the nanogap. Our device therefore benefits both the optical and electrical design by improving light-matter interaction for the former and reducing the volume and separation between the electrical contacts for the latter, resulting in an efficient mixed-mode device.

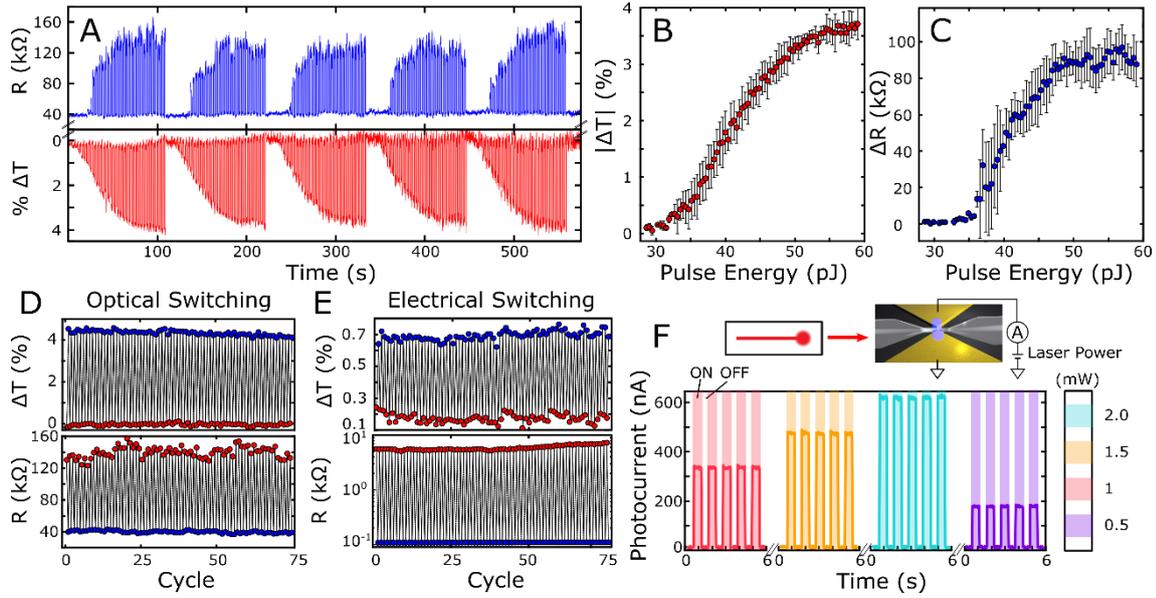

**Figure 3:** Multilevel operation and cyclability of mixed-mode device. **a)** Five consecutive cycles of multilevel operation with a fixed write pulse and a linearly increasing erase pulse energy (8 ns pulse width, 80 erase pulses per cycle). The variation in resistance is much greater than in optical transmission due to the stochastic nature of crystal domain growth within the nanogap. Cumulative plots of change in **b)** resistance and **c)** optical transmission for multilevel traces shown in **a)**. Cyclability plots of both the electrical resistance and optical transmission during multiple **d)** optical and **e)** electrical write and erase cycles. **f)** Photoconductive effect of device in amorphous state. Photocurrent is observed when the device is biased with a DC voltage (100 mV) due to filling of trap states in the GST. Sensitivity is enhanced due to the strong light-matter interaction in the plasmonic nanogap.

As the optical transmission and electrical conductance are both dependent on the fractional volume of crystalline domains, both are dependent on the energy of the write and erase pulses. Fig. 3a-c shows the dependence of both the optical transmission and electrical resistance of our device for various optical erase pulse energies (fixed 8 ns pulse width). Between each erase pulse, a fixed piecewise write pulse (1.3 nJ total energy and 408 ns in duration) is used to return the device to the crystalline state. We note that due to the stochastic nature of the formation of amorphous and crystalline domains within the memory cell, the resistance trace of Fig. 3a shows more variation than the read-out of the optical probe during optical switching cycles. On the other hand, the variation of the optical transmission is largely limited by the SNR of our optical read-out which can be addressed by increasing the coupling efficiency between the waveguide and plasmonic nanogap, which based on previous work could be improved by a factor of 3 (*29*). We attribute this to the different mechanisms between electrical conductance and optical transmission in our device. While the change in optical transmission depends on the overlap between the optical mode and the fractional volume of crystalline versus amorphous domains, a change in electrical conductance requires a continuous path of crystalline domains across the device to be formed—similar to a percolation network (*34*). This results in a dependence on both the factional volume of crystalline GST *and* the position of those domains which results in greater variation of the device resistance as seen in Fig. 3a. Because of this, the electrical domain has both a higher switching energy threshold and saturates at a lower erase pulse energy than the optical transmission read-out as shown in Fig. 3b and c. The lower saturation threshold is due to the fact that once the circuit is fully broken by an erase pulse, greater pulse energies will not affect the overall conductance of the device.

Our devices show good cyclability in both the electrical resistance and transmission for optically- and electrically-induced switching between the amorphous and crystalline states

(see Fig. 3d and e). The optical and electrical properties of GST and related phase-change chalcogenides have both been demonstrated commercially to be robust for over $10^{12}$ write-erase cycles (*9*, *35*). The cyclability of these phase-change chalcogenides, combined with a storage life-time of over 10 years at room temperature (*36*), make our approach highly promising for future integrated electro-optical storage. Table 1 provides a comparison of our work with other non-volatile photonic memories published to date. Not only do our results compare favourably with the literature in terms of active area and minimum switching energy, but we also show full optical and electrical programming and read-out for the first time in an integrated device.

| Device | Active Area (µm$^2$) | Min. Switch Energy (pJ) | Mixed-Mode Programming? | Mixed-Mode Read-out? | Non-Volatile? | Ref. |
|---|---|---|---|---|---|---|
| VO$_2$ on waveguide | 2.0 × 4.0 | 1.4×10$^3$ | No | Yes | No | (*22*) |
| VO$_2$ on waveguide | 0.35 × 0.5 | 9×10$^2$ | Yes | Yes | No | (*23*) |
| GeTe-NW on waveguide | 0.25 × 1.0 | 8×10$^3$ | No | Yes | Yes | (*24*) |
| GST on SPP waveguide | 0.5 × 2.0 | 6.9×10$^3$ | Yes | No | Yes | (*25*) |
| GST on waveguide | 1.0 × 1.3 | 67 ± 3 | No | No | Yes | (*30*) |
| GST plasmonic nanogap | 0.05 × 0.05 | 16 ± 2 | Yes | Yes | Yes | this work |

**Table 1:** Comparison of various mixed-mode, integrated photonic memory cells. Our work not only significantly reduces the active area of the device, but is also the first to allow full write-erase programming and read-out in both the optical and electrical domains.

**Conclusion**

We have demonstrated the first non-volatile nanoscale electro-optic device which enables both electrical and optical programming and read-out using the synergetic combination of phase change materials and nanoplasmonics. This is an unprecedented demonstration of an integrated, reversible, and non-volatile phase-change memory cell which fully bridges the gap between electro-optic mixed-mode operations. This was enabled by employing a plasmonic design which simultaneously reduces the footprint of the device, enhances light-matter interaction, and reduces the separation between electrical contacts, creating a compact and highly sensitive device. Our approach also enables a direct comparison of both optical and electrical read and write operations in a single device, demonstrating the relative merits and limitations of both. Importantly, the non-volatile nature our platform, provides an exciting outlook in the development of switchable and reconfigurable metadevices by means of optical or electrical stimuli; enabling novel approaches to switchable metamaterial-based optical components (*37–39*).We anticipate that a plethora of novel devices and platforms should arise in the coming years, which will capitalize on the bridge between the electrical and photonic domains that are demonstrated herein. These devices potentially herald true device-level integration of hybrid optoelectronic computing platforms with in-memory computing and multilevel data storage which is readily applicable to this work (*40*, *41*).

**Author Contributions**

All authors contributed substantially. NF and NY led the design, fabrication and testing of the devices. JT, XL, JS and ZC aided the modelling, characterization and the experimentation. CDW, WHPP and HB led the research, with NF, NY and HB drafting the manuscript.

**Acknowledgements**

This research was supported by EPSRC via grants EP/J018694/1, EP/M015173/1 and EP/M015130/1 in the UK; the Deutsche Forschungsgemeinschaft (DFG) grant PE 1832/2-1 in Germany; the European Research Council grant 682675; and from the European Union's Horizon 2020 research and innovation programme under grant agreement No 780848 (Fun-COMP project). HB thanks A. Ne for stimulating conversations. All authors thank the collaborative nature of European science for allowing this work to be carried out.